\documentstyle[12pt,psfig,epsf]{article}
\newcommand{\be}{\begin{equation}}
\newcommand{\qee}{\end{equation}}
\begin{document}
\begin{titlepage}
\title{
{\bf Gonihedric String Equation II}
}
{\bf
\author{ 
G.K.Savvidy\\
National Research Center "Demokritos",\\
Ag. Paraskevi, GR-15310 Athens, Greece \\
e-mail: savvidy@argo.nrcps.ariadne-t.gr\\
Tel. (301) 6518770. Fax. (301) 6511215
}
}
\date{}
\maketitle
\vspace{2cm} 
\begin{abstract}
\noindent

Arguing  that the equation for the gonihedric string should have 
a generalized Dirac form,  we found a new equation which corresponds
to a symmetric solution of the Majorana commutation relations and has 
non-Jacobian form. The corresponding generalized gamma-matrices are 
anticommuting and guarantee unitarity at all orders of $v/c$. The 
previous solution was in a Jacobian form and admits unitarity  
at zero order. Explicit formulas for the mass spectrum lead to nonzero 
string tension $M^{2}_{j} \geq M^{2}(j+1)^{2}$. 
The equation does not admit tachyonic solutions, but still has unwanted 
ghost solutions. We discuss also new dual transformation of the Dirac equation
and of the proposed generalizations.

\vspace{1cm} 
MCS codes: 81T13, 81T30, 83E30, 81R20

\vspace{1cm}

\end{abstract}
\thispagestyle{empty}
\end{titlepage}
\pagestyle{empty}

There is some experimental and theoretical evidence for the existence of a
string theory in four dimensions which may describe strong interactions and 
represent the solution of QCD \cite{nilsen}.

One of the possible candidates for that
purpose is the gonihedric string which has been defined as a model of 
random surfaces with an action which is proportional to the linear size 
of the surface \cite{sav2}
\be
A(M) = m\sum_{<ij>} \lambda_{ij}
\cdot \Theta(\alpha_{ij}),~~~~~~~ \Theta(\alpha)= \vert 
\pi - \alpha \vert^{\varsigma} , \label{action}
\qee
where $\lambda_{ij}$ is the length of the edge $<ij>$ of the 
triangulated surface $M$ and $\alpha_{ij}$ is the dihedral angle 
between two neighbouring triangles of $M$ sharing a common edge $<ij>$.
\footnote{The angular factor $\Theta$ and the idex $\varsigma$ 
define the rigidity 
of the random surfaces \cite{sav2} and the convergence of the partition 
function.  As it was proved in \cite{sav2}, the partition function 
$Z_{T}(\beta)$ for the given triangulation T is convergent when 
the parameter $\varsigma$ is in the interval $ 0< \varsigma \leq 1$. 
In \cite{durhuus} it has been proved that the full partition function
$Z(\beta) = \sum_{\{ T \}} Z_{T}(\beta)$ (the sum is over all triangulations 
$\{ T \}$)  is divergent for  $\frac{d-2}{d} < \varsigma \leq 1$, where d is the 
dimension of the spacetime. In \cite{sav2} it was demonstrated that for 
$0 <  \varsigma \leq \frac{d-2}{d}$ the $Z(\beta)$ is convergent and the scaling 
limit should be taken exactly at the point $ \varsigma = \frac{d-2}{d}$ so 
that the string tension (2) is generated. In addition to the formulation of 
the theory in the continuum space 
the system allows an equivalent representation on Euclidean lattices 
where a surface is associated with a collection of plaquettes 
\cite{wegner} and it has been 
proved  that the enthropy has exponential behaviour and not factorial.
The Monte Carlo 
simulations \cite{cut} demonstrate that the gonihedric 
system undergoes the 
second order phase transition and the string tension is generated 
by quantum fluctuations, as it was  expected theoretically 
(\ref{tension}).}
The model has a number of properties which make it very close to the Feynman
path integral for a point-like relativistic particle. 
In the limit when the surface degenerates into a single 
world line the action becomes proportional to the length of the 
path and the classical equation of motion for the gonihedric string
is reduced to the classical equation of motion for a free relativistic particle. 
At the classical level the string tension is equal to zero and, 
as it was demonstrated in \cite{sav2}, quantum fluctuations generate the 
nonzero string tension 
\be
\sigma_{quantum}= \frac{d}{a^2}~(1 - ln \frac{d}{\beta}) , \label{tension}
\qee
where $d$ is the dimension of the spacetime, $\beta$ is the coupling constant,
$a$ is the scaling parameter and $\varsigma = (d-2)/d$ in 
(\ref{action}).

It is natural therefore to ask what type of equation may describe this string
theory in the continuum limit.
The aim of the article \cite{sav13} was to suggest a possible answer to this
question. The analysis of the 
transfer matrix shows \cite{sav13} that the desired equation 
should describe propagation of fermionic degrees of freedom destributed 
over the space contour. When this contour shrinks to a point, the equation 
should describe propagation of a free Dirac fermion. Thus each
particle in this theory should be viewed as a state of a 
complex fermionic system and the system should have a point-particle 
limit when there is no excitation of the internal motion. In the given case 
this restriction should be 
understood as a principle according to which  the infinite sequence of 
particles should contain the spin one-half fermion and the 
equation should has the Dirac form  \cite{sav13}

\be
\{~i~\Gamma_{\mu}~\partial^{\mu}~~-~~M~\}~~\Psi~~~=0 . \label{stringeq}
\qee
The invariance of this equation under Lorentz transformations
$x'_{\mu} =\Lambda_{\mu}^{~\nu}~ x_{\nu},~~~\Psi'(x')=
\Theta(\Lambda)~ \Psi(x)$
leads to the following equation for the gamma matrices \cite{majorana,dirac1}
\be
\Gamma_{\nu} = \Lambda_{\nu}^{~\mu}~\Theta~\Gamma_{\mu}~\Theta^{-1}.
\qee
If we use the infinitesimal form of Lorentz transformations
$\Lambda_{\mu \nu}= \eta_{\mu \nu} + \epsilon_{\mu \nu},~~~\Theta =
1 + \frac{1}{2}\epsilon_{\mu \nu}~I^{\mu \nu}$
it follows that gamma matrices should satisfy the Majorana commutation relation
\cite{majorana}
\be
[\Gamma_{\mu},~I_{\lambda \rho}] = \eta_{\mu \lambda}~\Gamma_{\rho}
- \eta_{\mu \rho}~\Gamma_{\lambda}  \label{mcr}
\qee
where $I_{\mu \nu}$ are the generators of the Lorentz algebra. These equations 
allow to find the $\Gamma_{\mu}$ matrices when the representation of the 
$I_{\mu \nu}$ is given\footnote{Ettore Majorana suggested this extension of 
the Dirac equation in 1932 \cite{majorana} by constructing an 
infinite-dimensional extension of the gamma matrices.
An alternative way to incorporate the internal motion 
into the Dirac equation 
was suggested by Pierre Ramond in 1971 \cite{ramond}. 
In his extension of the Dirac equation the internal motion 
is incorporated through the construction of operator-valued gamma matrices. 
In both cases one can see effectively an extensions of Dirac gamma
matrices into the infinite-dimensional case.
For our purposes we shall follow Majorana's approach to incorporate the 
internal motion in the form of an infinite-dimensional wave equation.}.
The original Majorana solution 
for $\Gamma_{\mu}$ matrices is infinite-dimensional (see equation (14) in 
\cite{majorana}) and the mass spectrum of the theory is equal to
\be
M_{j}=\frac{M}{j+1/2},               \label{majorana}
\qee
where $j=1/2,3/2,5/2,....$ in the fermion case and $j=0,1,2,....$ in the 
boson case.
The main problems of Majorana theory are the decreasing mass spectrum 
(\ref{majorana}), 
absence of antiparticles and troublesome tachyonic solutions - the problems 
common to high spin theories \cite{grodsky}. 

Nevertheless in \cite{sav13}   
the Majorana theory has been interpreted as a natural way to incorporate 
additional degrees of freedom into the relativistic Dirac equation.   
Unlike Majorana the authors consider the infinite sequence of high-
dimensional representations of the Lorentz group with  nonzero Casimir
operators $(\vec{a}\cdot\vec{b})$ and $(\vec{a}^2 -\vec{b}^2)$. 
These representations $(j_{0},\lambda)$ and 
their adjoint $(j_{0},-\lambda)$ are enumerated by the 
index $r=j_{0} +1/2$, where $r=1,...,N$ and $j_{0}=1/2,3/2,...$~ is 
the lower spin in the representation $(j_{0},\lambda)$, thus 
$j=j_{0},j_{0}+1,...$. We took the free complex parameter $\lambda$ in
the real interval $-3/2 \leq \lambda \leq 3/2 $ in order to have $real$
matrix elements for the Lorentz boosts operator $\vec{b}$ (see 
(\ref{19}) and (\ref{lamb})). These representations are infinite-dimensional
except of the case $j_{0}=1/2,~\lambda = \pm 3/2$. At the same time their
{\it dual representations} were also used \cite{sav13}. 
The dual transformation
$(j_{0};\lambda)  \rightarrow (\lambda;j_{0})$, defined in \cite{sav13}, 
leads to a subsequent restriction on a free parameter $\lambda$ and requires 
$\lambda =1/2$ so that the dual representations become 
finite-dimensional $(1/2, \pm (1/2 + r))$. The corresponding 
$\Sigma^{dual}_{2}$-equation is not in contradiction with no-go theorem of
\cite{grodsky}, because dual representations are finite-dimensional.

In the present article we found a new $\Sigma \Delta$-equation which corresponds
to a symmetric solution of the Majorana commutation relations and has non-Jacobian
form. It is based on the same dual representations $(1/2, \pm (1/2 + r))$ of the 
Lorentz algebra and is a natural extension of the previous 
$\Sigma^{dual}_{2}$-equation of \cite{sav13}. 
The corresponding gamma-matrices are anticommuting 
\be
 \{ \Gamma_{\mu}, \Gamma_{\nu} \}  = 2~g_{\mu\nu} \Gamma^{2}_{0},
\qee
and guarantee unitarity at all orders of $v/c$. The $\Sigma^{dual}_{2}$-
equation admits unitarity at zero order.

For the completeness we shall review the logical and analytical steps which 
lead to $\Sigma^{dual}_{2}$-equation \cite{sav13} and then will derive 
the new equation which has anticommuting gamma-matrices.
In terms of $SO(3)$ generators~ $\vec{a}$~ and Lorentz 
boosts~ $\vec{b}$~~~($a_{x}
= iI^{23} ~~~ a_{y} = iI^{31}~~~a_{z} = iI^{12}~~~b_{x} 
 = iI^{10} ~~~ b_{y}=  iI^{20}~~~b_{z} = iI^{30}$)~~the
algebra of the $SO(3,1)$ generators 
can be rewritten as \cite{majorana,heisenberg} (we use Majorana's notations)
\be
[a_{x},  a_{y}]= ia_{z}~~~~[a_{x},  b_{y}]= ib_{z}~~~~[b_{x},  b_{y}]= 
-ia_{z} . \label{alge3} 
\qee
The irreducible representations $R^{(j)}$ of the $SO(3)$ subalgebra 
(\ref{alge3}) are 
\begin{eqnarray}
<j,m\vert ~a_z ~\vert j,m> =m \nonumber\\
<j,m\vert ~a_{+} ~\vert j,m-1> = \sqrt{(j+m)(j-m+1)} \nonumber\\
<j,m\vert ~ a_{-} ~\vert j,m+1> = \sqrt{(j+m+1)(j-m)}, \label{matr}
\end{eqnarray}
where $m=-j,...,+j$,~~~the dimension of  $R^{(j)}$ is~~ $2j+1$~~ and 
$j =0,~ 1/2,~1,~3/2,~...$ The representation $\Theta = (j_{0};\lambda)$ 
of the Lorentz algebra can be parameterized as \cite{heisenberg,majorana}
\begin{eqnarray}
<j,m \vert ~b_{z} ~\vert j,m> = \lambda_{j} \cdot m \label{7}\nonumber \\
<j-1,m \vert~ b_{z}~ \vert j,m> =\varsigma_{j}\cdot \sqrt{(j^2-m^2)} \nonumber\\
<j,m\vert~ b_{z}~ \vert j-1,m> = \varsigma_{j} \cdot \sqrt{(j^2-m^2)} \label{19}
\end{eqnarray}
plus similar formulas for the $b_{x}$ and $b_{y}$ generators.
The amplitudes $\lambda_{j}$ describe {\it diagonal} transitions 
inside the $SO(3)$ multiplet $R^{(j)}$, while  $\varsigma_{j}$ describe
{\it nondiagonal} transitions between $SO(3)$ multiplets which form the 
representation $\Theta$ of $SO(3,1)$. Thus  
$
\Theta(j_{0},\lambda) = \oplus \sum^{\infty}_{j=j_{0}}~R^{(j)},$
where $j_{0}$ defines the lower spin in the representation 
and $\lambda$ is a free complex parameter.
The amplitudes $\lambda_{j}$ and $\varsigma_{j}$ can 
be found from the commutation relations (\ref{alge3}) \cite{heisenberg,majorana}
\be
\lambda_{j}~~=~~ i\frac{j_{0}~\lambda}{j(j+1)},~~~~~~~~~\varsigma^{2}_{j} = 
\frac{(~j^{2} -j^{2}_{0}~)~(~j^2 -\lambda^{2})}
{j^2~(~4j^2-1~)} ,                       \label{lamb}
\qee
where $\lambda$ appears as an essential dynamical parameter which 
cannot be determined solely from the kinematics of the Lorentz group
\footnote{The representation is finite-dimensional if $\lambda=j_{0} +r$, 
$r=1,2,3,...$, as it is easy to see from (\ref{19}) and (\ref{lamb})).
The representations used in the 
Dirac equation are $(1/2,-3/2)$ and $(1/2,3/2)$ and
in the Majorana equation they are $(0,1/2)$ in the boson case and $(1/2,0)$ in
the fermion case. The infinite-dimensional Majorana representation $(1/2,0)$ 
contains $j=1/2,3/2,...$ 
multiplets of the $SO(3)$ while $(0,1/2)$ contains $j=0,1,2,...$ multiplets.}. 
The adjoint representation is defined as
$\dot{\Theta} = (j_{0};-\lambda)$. We shall consider the case 
$\Theta_{r} =(r-1/2,\lambda)$   and  
$-3/2 \leq \lambda \leq 3/2 $ to have $\varsigma_{j}$ real for all values of 
$r=1,2,..$ The Casimir operators $(\vec{a}\cdot \vec{b})$ and $~(\vec{a}^2 - 
\vec{b}^2)~$ for the representation $\Theta_{r}$ are equal 
correspondingly  to 
$<j,m\vert ~\vec{a} \cdot \vec{b}~ \vert j,m>~ =~ i~\lambda~ (r -1/2),~~~
<j,m\vert ~(\vec{a}^2 -\vec{b}^2)~ \vert j,m> ~= ~(r-1/2)^2 + 
\lambda^2 -1.$ As it is easy to see from these formulas the Casimir operator 
$(\vec{a}\cdot \vec{b})$ is nonzero only if $\lambda \neq 0$.

The Majorana commutation relation (\ref{mcr}) together with the last 
equations allow to find 
$\Gamma_{\mu}$ matrices when a representation $\Theta$ of the Lorentz algebra 
$I_{\mu\nu}$ is given \cite{majorana}. 
Because $\Gamma_{0}$ commutes with spatial rotations $\vec{a}$ (see (\ref{mcr}))
it should have the form
\be
<j,m\vert~\Gamma^{rr'}_{0}~\vert j'm'> = \gamma^{r r'}_{j} \cdot
\delta_{j j'}\cdot \delta_{m m'}~~~~~~~r,r' = \dot{N},...,\dot{1},
1,...,N   \label{solution1}
\qee
where we consider $N$ pairs of adjoint representations 
$\Theta=(\Theta_{\dot{N}},\cdots,
\Theta_{\dot{1}},\Theta_{1},\cdots, \Theta_{N})$ with $j_{0}= 
1/2,...,N-1/2$.
Thus $\gamma^{rr'}$ is $2N \times 2N$ matrix which 
should satisfy the equation for $\Gamma_{0}$ which follows from (\ref{mcr})
\cite{majorana}
\be
\Gamma_{0}b^{2}_{z}~ -~ 2~b_{z}\Gamma_{0}b_{z}~  + ~b^{2}_{z}\Gamma_{0}~ = ~-
\Gamma_{0}. \label{funeq}
\qee 
In \cite{sav13} the authors were searching the solution
of the above equation in the form of Jacoby matrices  
\be
\gamma_{j} = \left( \begin{array}{c}~~~0~~~~,~~~~~\gamma_{j}^{NN-1}
,~~~~~~~~~~~~~~~~~~~~~~~~~~~~~~~~~~~~~~\\
\gamma_{j}^{N-1N},~~~0~~~,~~~\gamma_{j}^{N-1N-2}
,~~~~~~~~~~~~~~~~~~~~~~~~~~~~~\\................................
..................\\..................................
\\~~~~~~~~~~~~~~~~~~~~~~~~~~~~\gamma_{j}^{\dot{N-1}\dot{N-2}}
,~~~0~~~,~~~~~\gamma_{j}^{\dot{N-1}\dot{N}}
\\~~~~~~~~~~~~~~~~~~~~~~~~~~~~~~~~~~~~~~~\gamma_{j}^{\dot{N}\dot{N-1}}
,~~~~0~~~~~\end{array} \right),~~~\Psi_{j}=\left( \begin{array}{c}
        \psi^{N}_{j}\\
        \psi^{N-1}_{j}\\
         .....\\
         .....\\
        \psi^{\dot{N}}_{j}\\
        \psi^{\dot{N-1}}_{j}
\end{array} \right). \label{ansatz}
\qee
It should be understood that 
$\vec{a}^{~rr'} = \delta^{rr'} \cdot \vec{a}~~~~~~~~\vec{b}^{~rr'} =
\delta^{rr'} \cdot \vec{b}^{~r}$ and the corresponding matrices $\vec{b}^{r}$ 
and $\vec{b}^{\dot{r}}$ are defined by (\ref{19}) and (\ref{lamb}). 
In the present work we found a new solution which has additional 
nonvanishing antidiagonal elements $\gamma^{r\dot{r}}_{j}$.

The equation (\ref{funeq}) together with
the matrix elements (\ref{solution1}),(\ref{ansatz}) and (\ref{19}),
(\ref{lamb}) completely define the problem. 
The solutions of the equation (\ref{funeq}) are defined up 
to a set of constant factors which are independent from $j$.
Indeed, because Jacoby matrices (\ref{ansatz}) have a very 
specific form, the original equation (\ref{funeq}) factorize into separate 
equations for every element $\gamma^{rr+1}_{j}$ of the Jacoby matrix 
and one can check that the solution has the form \cite{sav13}
\be
\gamma^{rr+1}_{j}~=~Const~\sqrt{(1-\frac{r^{2}}{N^{2}})}~\cdot
\sqrt{(\frac{j^2 +j}{4r^2 -1} -\frac{1}{4})} \label{jacsolu}
\qee
and has therefore $(4N-2)$ {\it j-independent free constant}. This 
freedom allows to impose necessary physical constraints on a solution 
requiring:   i) correct behaviour of the spectrum, 
ii) Hermitian property of the system, iii) reality and  the positivity of 
the current density 
matrix $\rho = \Omega \Gamma_{0}$. For that one should study the 
spectral properties of the matrices of infinite size with matrix elements 
$\gamma^{rr+1}_{j}$ and $\gamma^{r\dot{r}}_{j}$ which have complicated "root" 
dependence. The first inspection of the solution (\ref{jacsolu}) 
simply shows that every element  
$\gamma^{rr+1}_{j}$ grows like $\approx j$ and in general  all eigenvalues 
$\epsilon_{j}$ will also grow with j. 
Therefore the mass spectrum $ M_{j} =  M/\epsilon_{j}$ will have Majorana-like
behaviour (6) $ M_{j} \approx  M/j $. To avoid this general behaviour 
of the spectrum one should carefully inspect eigenvalues of the matrix 
$\Gamma_{0}$ for small values of $N$ and then for arbitrary N \cite{sav13}. The 
parameter N plays the role of a natural regularization.
The $B-H-\Sigma-\Sigma_{1}-\Sigma_{2}$-solutions which appear 
(see \cite{sav13} and below) have exeptional 
behaviour: half of the eigenvalues of the spectrum are increasing and 
the other half are decreasing. One can achieve this exeptional 
behaviour of the solution by tuning the free constants in the  
general solution (\ref{jacsolu}). However  these solutions have not been 
accepted \cite{sav13} because half of the eigenvalues produce a mass 
spectrum which has an accumulation point at zero mass. This phenomena can 
be understud on the example of the Dirac equation. For that let us define
the {\it dual representation} as 
$\Theta = (j_{0};\lambda)  \rightarrow (\lambda;j_{0})
= \Theta^{dual}$. From formulas (\ref{lamb}) and (\ref{19}) it is easy to see 
that 
representations $(j_{0},\lambda)$ and $(-j_{0},-\lambda)$ should be considered
as identical. Therefore  the dual transformation of the adjoint 
representation $\dot{\Theta} = (j_{0};-\lambda)$ which is defined as 
$(-\lambda;j_{0})$ is identical with $(\lambda;-j_{0})$, thus 
$\dot{\Theta} = (j_{0};-\lambda)  \longleftrightarrow (\lambda;-j_{0})
= \dot{\Theta}^{dual}$. For the dual representations $\Theta$ 
and $\Theta^{dual}$ the matrix elements of Lorentz
generators $I_{\mu\nu}$ are  precisely the same, the only difference 
between them is that the lower spin is equal to $j_{0}$ for the reperesentation 
$\Theta$ and is equal to $\lambda$ for its dual $\Theta^{dual}$ 
( see formulas (\ref{lamb}) and (\ref{19}) ).
Therefore any solution $\Gamma_{\mu}$
of the Majorana commutation relations (\ref{mcr}) for  $\Theta$ can be 
translated into the corresponding solution $\Gamma^{dual}_{\mu}$
for $\Theta^{dual}$ by exchanging $j_{0}$ for $\lambda$ \cite{sav13}. This
symmetry transformation imposes constraints on the free parameter 
$\lambda$, so that it should be {\it integer or half-integer}.

The dual transformation
of the Dirac representations  $(1/2,-3/2)$ and $(1/2,3/2)$ would be 
infinite-dimensional  $(3/2,-1/2)$ and $(3/2,1/2)$ with 
$j=3/2,5/2,...$ and the corresponding solution $\Gamma^{dual}_{0}$ 
has the form 
$\gamma^{1~\dot{1}}_{j} = \gamma^{\dot{1}~1}_{j} = j+1/2$ with the 
following mass spectrum 
\be
M^{Dirac~dual}_{j}=\frac{M}{j+1/2},~~~~~~~~ j=3/2,5/2,....    \label{10}     
\qee
This Majorana-like mass spectrum is dual to the physical spectrum 
of the Dirac equation
\be
M^{Dirac}_{j}= M,~~~~~~~~ j=1/2.       \label{11}  
\qee
The dual equation is simply unphysical, but we have to admit   
that the whole decreasing mass spectrum of the dual equation 
{\it corresponds or is dual} to a physical Dirac fermion. 
From this point of view we have to ask  about 
physical properties of the equations which are dual to "unphysical" ones 
$B-H-\Sigma-\Sigma_{1}-\Sigma_{2}$. The dual 
transformation completely improves the decreasing mass spectrum 
of these equations \cite{sav13} as it take place  in (\ref{10}) and (\ref{11}).
Indeed the last $\Sigma^{dual}_{2}$-equation has the spectrum of particle and
antiparticles of increasing half-integer spin lying on quasilinear trajectories.
The $\Sigma^{dual}_{2}$-equation admits unitarity only at zero order of $v/c$.

Before going on to review $B-H-\Sigma-\Sigma_{1}-\Sigma_{2}$-solutions and 
present the new $\Sigma\Delta$ equation let us introduce the invariant 
scalar product~$
<~\Theta ~\Psi_1~\vert~ \Theta ~\Psi_2~>~=~<~\Psi_1~\vert~ \Psi_2~> 
$, where $\Theta = 1 + \frac{1}{2}\epsilon_{\mu \nu}~I^{\mu \nu}$ and  
the matrix $\Omega$ is  defined as
$
<~\Psi_1~\vert~ \Psi_2~> = \bar{\Psi}_{1}~\Psi_{2} =\Psi^{+}_{1}~\Omega~\Psi_{2}
 = \Psi^{*~r}_{1~jm}~\Omega^{rr'}_{jm~j'm'}~\Psi^{r'}_{2~j'm'}
$ with the properties
\be
\Omega~a_{k} = a_{k}~\Omega~~~~
\Omega~b_{k} = b^{+}_{k}~\Omega ~~~~
\Omega = \Omega^{+}.     \label{gode} 
\qee
From the first relation it follows that 
$\Omega = \omega^{rr'}_{j}\cdot\delta_{jj'}\cdot\delta_{mm'}$
and from the last two equations, for our choice of the representation $\Theta$
and for a real $\lambda$ in the interval $-3/2 \leq \lambda \leq 3/2 $, that
$
\omega^{r\dot{r}}_{j}=\omega^{\dot{r}r}_{j}~=~1~~~\omega^{2}_{j}=1,
$ thus $\Omega$ is an antidiagonal matrix. The conserved current density 
is equal to
$J_{\mu} = \bar{\Psi}~\Gamma_{\mu}~\Psi,~~\partial^{\mu}~J_{\mu}=0.$
The current density $J_{0}$ should be {\it real and positive definite},
which is equivalent to the requirement that 
\be
\Gamma^{+}_{\mu}~\Omega =\Omega~\Gamma_{\mu}, \label{holdc}
\qee
and to the positivity of the eigenvalues of the matrix 
$\rho = \Omega~\Gamma_{0}$.

The basic solution ( {\it B-solution}) of the equation (\ref{funeq}) for the 
$\Gamma_{0}$ has the form (\ref{ansatz}), (\ref{jacsolu}) with all set of constant 
factors equal to $Const=i$ \cite{sav13}
\be
\gamma^{r+1~r}_{j}=\gamma^{r~r+1}_{j} =
\gamma^{\dot{r+1}~\dot{r}}_{j}= \gamma^{\dot{r}~\dot{r+1}}_{j} = 
i~ \sqrt{(1-\frac{r^2}{N^2})~(\frac{j^2 + j}
{4r^2-1} -\frac{1}{4})}~~~~~j\geq r+1/2                   \label{solution} 
\qee
and 
$
\gamma^{1~\dot{1}}_{j} = \gamma^{\dot{1}~1}_{j} = j+1/2
$,
where $r=1,...,N-1$. 
These matrices grow in size with $j$ until $j=N - 1/2$, for greater  $j$ 
the size of the matrix $\gamma_{j}$ remains the same and is equal to
$2N \times 2N$. The number of states with angular momentum $j$  grows as
$j+1/2$ and  this takes place up to spin $j=N-1/2$. For higher spins 
$j \geq N-1/2$ the number of states remains constant and is equal to $N$.
The positive eigenvalues $\epsilon_{j}$ can  now be found  \cite{sav13}
$$
        1~~~~~~~~~~~~~~~~~~~~~~~~~~~~~j=1/2  \nonumber
$$
$$
1-1/N~~~~~~~~1+1/N~~~~~~~~~~~~~~~~~~~~~~~~~~~~~~~~j=3/2      \nonumber
$$
.~.~.~.~.~.~.~.~.~.~.~.~.~.~.~.~.~.~.~.~.~.~.~.~.~.~.~.~.~.~.
\be
\cdots~~,1 + (j-5/2)/N,~~~1+(j-1/2)/N~~~~~~~~~j\geq N-1/2
\qee
The last formulas show that the coefficient of 
proportionality behind $j$ drops N times compared with the one in 
the Majorana solution $\epsilon_{j}=j+1/2$ in (6)
and now  many eigenvalues are less 
than unity and the corresponding masses $M_{j}= M/\epsilon_{j}$
are bigger than the ground state mass M. This actually means 
that by increasing the number of representations in $\Theta=
(\Theta_{\dot{N}},\cdots,\Theta_{\dot{1}},\Theta_{1},\cdots, \Theta_{N})$ 
one can slow down the growth of the eigenvalues. To have  
the mass spectrum  bounded from below one should have  spectrum with 
all eigenvalues $\epsilon_{j}$ less than unity.
In the limit $N \rightarrow \infty$ the B-solution (\ref{solution})
is being reduced to the form
\be
\gamma^{r+1~r}_{j}=\gamma^{r~r+1}_{j} =
\gamma^{\dot{r+1}~\dot{r}}_{j}= \gamma^{\dot{r}~\dot{r+1}}_{j} = 
i~ \sqrt{(\frac{j^2 + j}
{4r^2-1} -\frac{1}{4})}~~~~~j\geq r+1/2  
\qee
and
$
\gamma^{1~\dot{1}}_{j} = \gamma^{\dot{1}~1}_{j} = j+1/2
$,
where $r=1,2,...$. As it is easy to see from the previous formulas, 
all eigenvalues $\epsilon_{j}$ 
tend to unity when the number of representations
$N \rightarrow \infty$.
The characteristic equation which is satisfied by the gamma matrix in 
this limit is
\be
(\gamma_{j}^2 -1)^{j+1/2} = 0 ~~~~~~~~~~j=1/2,~3/2,~,5/2,\cdots \label{det1}
\qee
with all eigenvalues $\epsilon_{j} = \pm 1$. Therefore all states have equal
masses $M_{j}= 1$ and the spectrum is bounded from below, but the Hamiltonian 
is not Hermitian ($\Gamma^{+}_{0} \neq \Gamma_{0}$) \footnote{The determinant
and the trace are equal to
$
Det~\gamma_{j} = \pm 1,~~Tr~\gamma^{2}_{j} = 2j +1,  
$
thus~$
\epsilon_{1}^2 \cdot ...\cdot\epsilon_{j+1/2}^2= 1,~~~\epsilon_{1}^2 + 
...+\epsilon_{j+1/2}^2 = j+1/2.
$}.
The matrix $\Omega~\Gamma_{0}$ has the characteristic equation
$
(\omega_{j}~\gamma_{j} - 1)^{2j + 1} = 0
$
with all eigenvalues equal to $\rho_{j} = +1$.
Thus the matrix $\Omega~\Gamma_{0}$ is positive definite and all its 
eigenvalues are equal to one, but the relations
$
\Omega~\Gamma_{0} \neq \Gamma^{+}_{0}~\Omega,~~\Gamma^{+}_{0} \neq
\Gamma_{0}
$
do not hold. What is crucial
here is that we can improve the B-solution without 
disturbing its determinant which is equal to one (\ref{det1}) 
$(Det\Gamma_{0} =1)$.  The last property of the determinat is  
necessary to keep in order that the spectrum will be symmertically 
distributed arround unity. 

The Hermitian solution ({\it H-solution}) of (\ref{funeq}) for $\Gamma_{0}$
can be found as a phase modification of the basic {\it B-solution} \cite{sav13}
(\ref{solution})
\be
\gamma^{r+1~r}_{j}=-\gamma^{r~r+1}_{j} =
-\gamma^{\dot{r+1}~\dot{r}}_{j}= \gamma^{\dot{r}~\dot{r+1}}_{j} =
i~ \sqrt{(\frac{j^2 + j}{4r^2-1} -\frac{1}{4})}~~~~~j\geq r+1/2 .  \label{herso}
\qee
These matrices are Hermitian $\Gamma^{+}_{0}=\Gamma_{0}$, but
the characteristic equations are more complicated now. 
These  polynomials $p(\epsilon)$ have reflective symmetry and are even
$p_{j}(\epsilon)~= ~\epsilon^{2j+1}~p_{j}
(1/\epsilon)$,~~$p_{j}(-\epsilon)~= ~p_{j}(\epsilon)$ therefore 
if $\epsilon_{j}$ is a solution then $1/\epsilon_{j}$,~$-\epsilon_{j}$
and $-1/\epsilon_{j}$ are also solutions \footnote{
Computing the traces and determinants of these matrices one can get 
the following general relation for the eigenvalues
$
\epsilon_{1}^2 \cdot ...\cdot\epsilon_{j+1/2}^2= 1,~~~\epsilon_{1}^2 + 
...+\epsilon_{j+1/2}^2 = j(2j+1).
$}.
The eigenvalues $\epsilon_{j}$ can be found \cite{sav13}
$$
~~~~~~~~~~~~~~~~~~~1~~~~~~~~~~~~~~~~~~~~~~~~~~~~~~~~~~~j=1/2       \nonumber    
$$
$$
~~~~~~~~~~~~~~~~~\sqrt{2}-1~~~~~~~~\sqrt{2}+1~~~~~~~~~~~~~~~~~~~~~~~~~~~~~~~~~~j=3/2
\nonumber
$$
\be
.~.~.~.~.~.~.~.~.~.~.~.~.~.~.~.~.~.~.~.~.~.~.~.~.~.~.~.~.~.~.~.~.~.~.~.\label{osp}
\qee
The changes in the phases of the matrix elements (\ref{herso}) 
result in a different behaviour of eigenvalues.
The half of the eigenvalues (decreasing eigenvalues), produce quasilinear 
trajectories with nonzero string tension
and the other half (increasing eigenvalues) affect the low spin states 
on  trajectories, so that smallest mass on a given trajectory  
tends to zero (see \cite{sav13}). The matrix $\Omega~\Gamma_{0}$ has again the 
characteristic equation
$
(\omega_{j}~\gamma_{j} - 1)^{2j + 1} = 0
$
and all eigenvalues are equal to one. Thus again the  
matrix $\Omega~\Gamma_{0}$ is positive definite 
because all eigenvalues are equal to one, but the important relation
$\Omega~\Gamma_{0} \neq \Gamma^{+}_{0}~\Omega$
does not hold. 
The  solution of (\ref{funeq}) for $\Gamma_{0}$ with both properties
$\Gamma^{+}_{0} = \Gamma_{0}$ and $\Omega~\Gamma_{0} = \Gamma^{+}_{0}~\Omega$
can be found by using the basic solutions (\ref{solution}) rewritten with
arbitrary phases of the matrix elements and then by requiring that $\Gamma_{0}$
should be Hermitian $\Gamma^{+}_{0} = \Gamma_{0}$ and should satisfy the 
relation $\Omega~\Gamma_{0} = \Gamma^{+}_{0}~\Omega$. This solution,  
$\Sigma$-$solution$, is symmetric and has the form \cite{sav13}
\be
\gamma^{r+1~r}_{j}=~\gamma^{r~r+1}_{j} =
\gamma^{\dot{r+1}~\dot{r}}_{j}= \gamma^{\dot{r}~\dot{r+1}}_{j} =
 \sqrt{(\frac{j^2 + j}{4r^2-1} -\frac{1}{4})}~~~~~j\geq r+1/2.   \label{realsym}
\qee
In this case the Hermitian matrix $\Gamma^{+}_{0} = \Gamma_{0}$ 
has the desired property 
$
\Gamma^{+}_{0}~\Omega = \Omega~\Gamma_{0}.
$
This  means that the current density is equal to $\rho =\Omega~\Gamma_{0}$.
In addition, all of the gamma matrices now have this property (\ref{holdc})
$
\Gamma^{+}_{k}~\Omega = \Omega~\Gamma_{k}~~k=x,y,z
$
which follows from the equation $\Gamma_{k}= i [b_{k}~\Gamma_{0}]$ (\ref{mcr}) 
and equation (\ref{gode}) $\Omega~b_{k}~ = b^{+}_{k}~\Omega$.

The characteristic equations 
and the spectrum (\ref{osp}) are the same for the Hermitian 
H-solution and symmetric $\Sigma$-solution, but 
the corresponding characteristic equations for the matrices $\rho_{j}$ are
different and the eigenvalues of the density matrix are not positive
definite any more 
$$
~~~~~~~~~~~~~~~~~~~~~~~~~1~~~~~~~~~~1~~~~~~~~~~~~~~~~~~~~~~~~~~~~~~~~~~~~~j=1/2        \nonumber
$$
$$
~~~~~~~~~~~~~~~~~1- \sqrt{2}~~~~~~~~\sqrt{2}+1~~~~~~~~~~~~~~~~~~~~~~~~~~~~~~~~j=3/2\nonumber
$$
\be
.~.~.~.~.~.~.~.~.~.~.~.~.~.~.~.~.~.~.~.~.~.~.~.~.~.~.~.~.~.~.~.~.~.~.~.~.~.~.~.~.~\label{ospc}
\qee
Both states with $j=1/2$ have positive norms, the $j=3/2$ level has two 
positive and two negative norm states, the $j=5/2$ has four positive and two 
negative norm states, and so on.
The positive norm physical states are lying on 
the quasilinear trajectories of different slope and  
the negative norm ghost states are also lying on the quasilinear 
trajectories \cite{sav13}. Thus the equation has the 
increasing mass spectrum, but the smallest mass on a given trajectory 
still tends to zero and in addition there are many ghost states
(see also bellow).

In the case when some of the transition amplitudes in (\ref{realsym}) are set 
to zero 
\be
\gamma^{\dot{1}~1}~ = ~ \gamma^{2~3}_{j}~=~\gamma^{4~5}_{j} 
= ....=0~~~~~\gamma^{1~\dot{1}}~=~\gamma^
{\dot{2}~\dot{3}}~=~\gamma^{\dot{4}~\dot{5}}~= ... = 0 \label{diaggammsq}
\qee
and all other elements of the $\Gamma_{0}$ matrix remain the same as in 
(\ref{realsym}) we have a new $\Sigma_{1}$-solution with the important 
property that $\Gamma^{2}_{0}$ is a diagonal matrix and that the antihermitian 
part of $\Gamma_{k}$ anticommutes with $\Gamma_{0}$. Thus in
this case we recover the nondiagonal 
part of the Dirac commutation relations for gamma matrices
$
\{ \Gamma_{0}, \tilde{\Gamma_{k}} \} =0~~~~~~~~k =x,y,z. 
$
For the solution (\ref{diaggammsq}) 
one can explicitly compute the mass spectrum and the slope of the trajectories 
\cite{sav13}
\be
M^{2}_{n}= \frac{2 M^{2}}{n}~ \frac{j^2 -(2n-1)j 
+n(n-1)}{j-(n-1)/2}~~n=1,2,... \label{strtension}
\qee
where $j = n+1/2, n+5/2, ....$.
The string tension $\sigma_{n} = 1/2\pi\alpha^{'}_{n}$
varies from one trajectory to another and is equal to
\be
2\pi \sigma_{n}  = \frac{1}{\alpha^{'}_{n}} = \frac{2M^2}{n}~~~~~~n=1,2,... 
\label{s}
\qee
Thus we have the string equation which has 
trajectories with different string tension and that trajectories 
with large $n$
are almost "free" because the string tension tends to zero.
The smallest mass   
on a given trajectory $n$ has spin $j = n+1/2$ and decreases as 
$
M^{2}_{n}(j=n+1/2)= \frac{3M^2}{n(n+3)} .
$
The other solution, $\Sigma_{2}$-solution, which shares the above properties 
of $\Sigma_{1}$-solution is (\ref{realsym}) with 
\be
\gamma^{1~2}_{j}~ = ~ \gamma^{3~4}_{j}~= ....
=0~~~~~~\gamma^{\dot{1}~\dot{2}}_{j}~=~\gamma^{\dot{3}~\dot{4}}_{j}~= ... = 0.
\label{delta2}
\qee
The difference between these last two solutions is that in the first case 
the lower spin is $j=3/2$ and in the second case it is $j=1/2$.
The unwanted property of all these solutions $\Sigma$, $\Sigma_{1}$ and 
$\Sigma_{2}$ is that the  smallest mass 
$M^{2}_{n}(min)$ tends to zero and the spectrum is not bounded from below. 
We have to remark also that both equations, $\Sigma_{1}$ and $\Sigma_{2}$, 
which correspond to  (\ref{diaggammsq}) 
and to (\ref{delta2}) have natural $constraints$ \cite{sav13}. 

The unwanted property of the $\Sigma$-solutions, that is the  
decreasing of the smallest mass
on a given trajectory, can be solved by dual transformation of the system 
\cite{sav13}. This {\it exact symmetry} transforms two solutions of the 
Majorana commutation relations one into another. Indeed, interchanging 
$j_{0}$ and $\lambda$ in the representation $\Theta = (j_{0},\lambda)$ 
does not affect the matrix elements of the Lorentz generators $I_{\mu\nu}$,
therefore a  solution $\gamma_{j}$ 
for $\Theta = (j_{0},\lambda)$ can be translated into 
a solution $\gamma^{dual}_{j}$ for the dual representation 
$\Theta^{dual}$ by exchanging $j_{0}$ for $\lambda$ and letting spin j 
to run in a different interval $j=\lambda, \lambda +1,...$
\footnote{The obvious consiquence of the dual transformation is that 
the free parameter $\lambda$ should be integer or half-integer.}. 
The dual transformation does not change the actual j dependence of  
$\gamma_{j}$, 
but what is important here is that despite the fact that the dual solution 
$\gamma^{dual}_{j}$ is  almost identical with $\gamma_{j}$
(we mean the j dependence),
the spectrum  essentially changes for the low spin statets and does not 
affect the high spin states. This is because the number of states 
with  spin j, which was equal to $j+1/2$ before the dual transformation   
becomes infinite now. 
Therefore the spin contents of the dual equation is different  
and the equation has different spectrum. 
The dual transformation  does not affect the higher spin states 
and thus
does not change the slope of the trajectories (\ref{s}), and 
has the commulative effect on  lower spin states keeping them bounded from
below.  

Indeed under the dual transformation  
$
\Theta = (j_{0};\lambda)  \rightarrow (\lambda;j_{0})
= \Theta^{dual}
$
the representation $\Theta=(\Theta_{\dot{N}},\cdots,
\Theta_{\dot{1}},\Theta_{1},\cdots, \Theta_{N})$  is transformed into its dual
$\Theta^{dual}=
.....(\lambda; -5/2)~~(\lambda; -3/2)~~(\lambda; -1/2)~~(\lambda; 
1/2)~~(\lambda;3/2)~~(\lambda; 5/2)....$ 
and we are lead to take $\lambda$ to be half-integer and to 
$\lambda = 1/2$ in order to have the Dirac
representation incorporated in $\Theta$ \footnote{These representations do 
not coinside with the ones in Ramond equation \cite{ramond}.}. 
The solution which is dual to 
$\Sigma_{2}$
(\ref{realsym}) and  (\ref{delta2})  is equal to \cite{sav13}
\be
\gamma^{r+1~r}_{j}=~\gamma^{r~r+1}_{j} =
\gamma^{\dot{r+1}~\dot{r}}_{j}= \gamma^{\dot{r}~\dot{r+1}}_{j} =
 \sqrt{(\frac{1}{4}-\frac{j^2 + j}{4r^2-1})}~~~~~r\geq j+3/2    \label{dualsym}
\qee
where $j=1/2,3/2,5/2,...$,~~$r=2,4,6,....$ and the rest of the elements are equal to zero
\be
\gamma^{\dot{1}~1}~ = ~ \gamma^{1~2}_{j}~=~\gamma^{3~4}_{j} 
= ....=0~~~~~\gamma^{1~\dot{1}
}~=~\gamma^{\dot{1}~\dot{2}}~=~\gamma^{\dot{3}~\dot{4}}~= 
... = 0.\label{dualsym1}
\qee
The Lorentz boost operators $\vec{b}$ are antihermitian in this case 
$b^{+}_{k} = -b_{k}$, 
because the amplitudes $\varsigma$ (\ref{lamb}) are pure imaginary and 
therefore the $\Gamma_{k}$ matrices are also antihermitian
$\Gamma^{+}_{k} = - \Gamma_{k}$.
The matrix $\Omega$ changes and is now equal to the parity operator $P$, the
relation $\Omega~\Gamma^{+}_{\mu}= \Gamma_{\mu}~\Omega$
remains valid. The diagonal part of $\Gamma_{k}$ anticommutes
with $\Gamma_{0}$
as it was before $\{ \Gamma_{0}, \tilde{\Gamma_{k}} \} =0~~k =x,y,z$.
The mass spectrum is equal to
\be
M^{2}_{n}= \frac{2M^2}{n}~\frac{(j+n)(j+n+1)}{j+(n+1)/2} 
\label{dualmassspectrum}
\qee
where $n=1,2,3,..$ and enumerates the trajectories. 
The lowest spin on a given trajectory is either $1/2$ or $3/2$ 
depending on n: if n is odd then 
$j_{min}=1/2$, if n is even $j_{min}=3/2$. This is an essential
new property
of the dual equation because now we have an infinite number of states 
with a given
spin $j$ instead of $j+1/2$.
The string tension is the same as in the dual system  (\ref{s}).  
The lower mass on a given trajectory $n$ is given by the formula $(j=1/2)$ is
$
M^{2}_{n}(j=1/2) = \frac{4M^2}{n}\frac{(2n+1)(2n+3)}{n+2} \rightarrow (4M)^2
$ 
and the spectrum is bounded from below by positive mass.

The last $\Sigma^{dual}_{2}$-equation 
has the property that only the diagonal matrix elements of the anticommutator
$\{ \Gamma_{0}, \Gamma_{z} \}$ are equal to zero
\be
<j,m,r \vert \{ \Gamma_{0}, \Gamma_{z} \} \vert r,j,m> = \gamma^{rr+1}_{j}
im(\lambda^{r+1}_{j}-\lambda^{r}_{j})\gamma^{r+1r}_{j} + im
(\lambda^{r}_{j}-\lambda^{r+1}_{j})\gamma^{rr+1}_{j} \gamma^{r+1r}_{j} 
=0, \label{anti}
\qee
and that nondiagonal elements are not equal to zero  
\be
<j-1,m,r \vert \{ \Gamma_{0}, \Gamma_{z} \} \vert r,j,m> = 
i\sqrt{j^2 - m^2}~\varsigma^{r}_{j}~[ (\gamma^{rr+1}_{j})^{2} - 
(\gamma^{rr+1}_{j-1})^{2}]. \label{antinon}
\qee
Let us search the solution of the Majorana commutation relation (\ref{funeq})
in the same $\Sigma^{dual}_{2}$-Jacoby form (\ref{ansatz}) but with an 
additional nonvanishing antidiagonal matrix elements $\gamma^{r~\dot{r}}_{j}$. 
The solution has the form 
\be
\gamma^{r~\dot{r}}_{j}=~\gamma^{\dot{r}~r}_{j}=~-\gamma^{r+1~\dot{r+1}}_{j} 
=~-\gamma^{\dot{r+1}~r+1}_{j}= \frac{j + 1/2}
{\sqrt{4r^2-1}}
\qee
where $j=1/2,3/2,5/2,...$,~~$r=2,4,6,....$ and $r \geq j +3/2$ and one can check
directly that $\Gamma_{0}$ is the solution of (\ref{funeq}). These additional 
matrix elements in $\Gamma_{0}$ will not change the diagonal matrix elements 
of the anticommutator (\ref{anti}) because 
\be
<j,m,r \vert \{ \Gamma_{0}, \Gamma_{z} \} \vert r,j,m> = \gamma^{r\dot{r}}_{j}
im(\lambda^{\dot{r}}_{j}-\lambda^{r}_{j})\gamma^{\dot{r}r}_{j} + im
(\lambda^{r}_{j}-\lambda^{\dot{r}}_{j})\gamma^{r\dot{r}}_{j}
 \gamma^{\dot{r}r}_{j} =0, \label{anti1}
\qee
and  will cancel nondiagonal matrix elements of  (\ref{antinon})
\be
<j-1,m,r \vert \{ \Gamma_{0}, \Gamma_{z} \} \vert r,j,m> = 
i\sqrt{j^2 - m^2}~\varsigma^{r}_{j}~[ (\gamma^{rr+1}_{j})^{2} - 
(\gamma^{rr+1}_{j-1})^{2}  +  (\gamma^{r\dot{r}}_{j})^{2}  
- (\gamma^{r\dot{r}}_{j-1})^{2}] =0. \label{antinon1}
\qee
One can check this fact also using the relation 
$\{ \Gamma_{0}~\Gamma_{z} \}=i~[b_{z}~\Gamma^{2}_{0}]$ which follows from
(\ref{mcr}).
Using the relations (\ref{mcr}) $\Gamma_{y} =-i~[\Gamma_{z}~a_{x}]$ and 
$[\Gamma_{0}~a_{x}]=0$ one can see that $\{ \Gamma_{0}, \Gamma_{y} \}=0$ and 
in the same way, that $\{ \Gamma_{0}, \Gamma_{x} \}=0$. Finally using the 
relation (\ref{mcr}) $\Gamma_{k} =-i~[\Gamma_{0}~b_{k}]$ one can prove by 
direct calculation
that  $\{ \Gamma_{k}, \Gamma_{l} \}=0$ for $k\neq l$ and then using 
(\ref{funeq}) and the fact that $[b_{k}~\Gamma^{2}_{0}]=0$  to prove that 
$\Gamma^{2}_{k}=-\Gamma^{2}_{0}$ therefore 
\be
\{ \Gamma_{\mu}, \Gamma_{\nu} \}~=~2 g_{\mu\nu}~ \Gamma^{2}_{0}, 
\qee
where $\Gamma^{2}_{0}$ is a diagonal martix. Now the theory is 
Hermitian in all orders of $v/c$.
The mass spectrum is highly degenerated and is given by the formula
\be
M^{2}_{j}= M^{2} \frac{4r^{2}-1}{4r^{2}}~~~~~~~~r \geq j +3/2.
\qee

New  mass terms $(\vec{a}\cdot\vec{b})~\Gamma_{5}$ and $(\vec{a}^2 -\vec{b}^2)$ 
can be added into the string equation (\ref{stringeq})
in order to increase the  string tension 
\be
\{~i~\Gamma_{\mu}~\partial^{\mu}~~-~~M~(\vec{a}
\cdot\vec{b})\Gamma_{5}~~-~~gM~(\vec{a}^2 -\vec{b}^2)~\}~~\Psi~~~=0  
\qee
where $(\vec{a}\cdot\vec{b})$ and $(\vec{a}^2 -\vec{b}^2)$ are the Casimir 
operators of the Lorentz algebra. 
The commutation relations which define the $\Gamma_{5}$ 
matrix are $\Gamma_{5}~a_{k} = a_{k}~\Gamma_{5},~~\Gamma_{5}~b_{k} = 
b_{k}~\Gamma_{5},~~\Gamma_{5}^2 = 1$, thus 
$\Gamma_{5} = \Gamma_{5~j}^{rr'}~\delta_{jj'}~\delta_{mm'}$ 
and 
$
\Gamma_{5~j}^{rr} = -\Gamma_{5~j}^{\dot{r}\dot{r}} = (-1)^{r+1}.
$~One can check that $\Gamma_{5}~\Gamma_{0} = - 
\Gamma_{0}~\Gamma_{5},~~\Gamma_{5}~\Gamma_{k} = -
\Gamma_{k}~\Gamma_{5},~~\Gamma_{5}~P = 
- P~\Gamma_{5},~\Gamma_{5}~\Omega = - \Omega~\Gamma_{5}$, where 
the parity operator P defined as~ $P~a_{k} = a_{k}~P,~P~b_{k} =
- b_{k}~P,~~P^2 = 1$. We have again $P = P^{rr'}_{j}~\delta_{jj'}~\delta_{mm'}$ 
and that
$
P^{r\dot{r}}_{j} = P^{\dot{r}r}_{j} = (-1)^{[j]}
$.
One can check that $P~\Gamma_{0} = \Gamma_{0}~P,~~P~\Gamma_{k} = - 
\Gamma_{k}~P,~~P~\Omega = \Omega~P$.
Thus the additional new mass matrix $(\vec{a}\cdot \vec{b}) ~ \Gamma_{5} $ 
is diagonal and is equal to~ $<j,m,r~\vert (\vec{a}\cdot \vec{b}) ~ \Gamma_{5}
\vert ~r,j,m>~ =~ <j,m,\dot{r}~\vert (\vec{a}\cdot \vec{b}) ~ \Gamma_{5}
 \vert ~\dot{r},j,m>,$~~$=i~\frac{1}{2}~(-1)^{r+1}(r -1/2).$

Including the $\Gamma_{5}$ mass term one can see that mass spectrum 
grows as $j^2$ and all trajectories acquire a nonzero slope
\be
M^{2}_{j}= \frac{M^2}{4}~\frac{4r^{2}-1}{r^{2}} (r-1/2)^2~~~~~~~r\geq j+3/2
\qee
where $ r=2,4,6,....$,~~$j=1/2,3/2,5/2,....$,
thus $M^{2}_{j} \geq M^{2}(j+1)^2$.
If we "turn on" the pure Casimir mass term $gM~(\vec{a}^2 -\vec{b}^2)$  the 
spectrum will grow as $j^{4}$
\be
M^{2}_{j}= (gM)^{2} ~\frac{4r^{2}-1}{r^{2}} (r-1/2)^4~~~~~~~r\geq j+3/2 ,
\qee
thus $M^{2}_{j} \geq (2gM)^{2}(j+1)^4$. In general case the formula is 
\be
M^{2}_{j}= \frac{M^2}{4}~\frac{4r^{2}-1}{r^{2}} (r-1/2)^{2}
(1+ 2g (r-1/2))^{2}~~~~~~~r\geq j+3/2.
\qee
Thus the  spectrum of the theory consists of 
particles and antiparticles of increasing half-integer spin 
lying on quasilinear trajectories of different slope.
It is difficult to say at the moment what is the 
physical reason for this nonperturbative behaviour. 
The equation is explicitly Lorentz invariant, but has unwanted ghost solutions.
The tachyonic solutions which appear in Majorana 
equation (see (20) in \cite{majorana}) do 
not show up here. This is because the nondiagonal transition amplitudes 
of the form
$<..j..\vert~\Gamma_{k}~\vert .. j \pm 1..>$ are small here and the 
diagonal amplitudes 
$<..j..\vert~\Gamma_{k}~\vert .. j.. >$ are large. The problem of ghost 
states is more 
subtle here 
and we shall analyze  the natural constraints appearing in the system  to 
ensure that they decouple from the physical space of states.

We will present the derivation of the above equation from the 
gonihedric string, which was formulated as a model of random surfaces, 
in a separate place. The equation has its own value independent of the
motivation advocated in this article. 

In conclusion I would like to acknowledge Konstantin Savvidy for stimulating 
discussions.
This work was supported in part by the EEC Grant no. ERBFMBICT972402.

\vfill

\begin{thebibliography}{99}

\bibitem{nilsen}H.B.Nielsen and P.Olesen. Phys.Lett. B32 (1970) 203\\
D.J.Gross and F.Wilczek. Phys.Rev.Lett.30(1973)1343\\
G. 't Hooft. Nucl.Phys. B72 (1974) 461; Nucl.Phys. B75 (1975) 461\\
A.Polyakov. Gauge fields and String. (Harwood Academic Publishers, 1987)\\
D.Gross and W.Taylor. Nucl.Phys. B400 (1993) 181 ; B403 (1993) 395\\
N. Seiberg and  E. Witten. Nucl.Phys.B426 (1994) 19

\bibitem{sav2}G.K. Savvidy and K.G. Savvidy. Mod.Phys.Lett. A8 (1993) 2963\\
R.V. Ambartzumian and et. all. Phys. Lett. B275 (1992) 99\\
G.K. Savvidy and K.G. Savvidy. Int. J. Mod. Phys. A8 (1993) 3993\\
G.K.Savvidy and R.Schneider. Comm.Math.Phys. 161 (1994) 283

\bibitem{sav13}G.K. Savvidy and K.G. Savvidy.  Gonihedric string equation I, 
hep-th/9711015 

\bibitem{durhuus}B.Durhuus and T.Jonsson. Phys.Lett. B297 (1992) 271


\bibitem{wegner}G.K.Savvidy and F.J.Wegner. Nucl.Phys.B413(1994)605\\
G.K. Savvidy and K.G. Savvidy. Phys.Lett. B324 (1994) 72\\
G.K. Savvidy and K.G. Savvidy. Phys.Lett. B337 (1994) 333;\\ 
Mod.Phys.Lett. A11 (1996) 1379\\
R. Pietig and F.J. Wegner. Nucl.Phys. B466 (1996) 513

\bibitem{cut}G.Koutsoumbas, G.K. Savvidy and K.G. Savvidy. 
Phys.Lett. B410 (1997) 241

\bibitem{majorana} E.Majorana. Teoria Relativistica di Particelle con Momento
Intrinseco Arbitrario. Nuovo Cimento 9 (1932) 335

\bibitem{grodsky}I.T.Grodsky and R.F.Streater. Phys.Rev.Lett. 20 (1968) 695\\
N.N.Bogoliubov and V.S.Vladimirov. N.D.V.S. Fiz.Mat.Nauk 3 (1958) 26\\
E.Abers,I.T.Grodsky and R.E.Norton.  Phys.Rev. 159 (1967) 1222\\
D.Stoynov and I.T.Todorov. J.Math.Phys. 9 (1968) 2146\\
R.Casalbuoni and G.Longhi. Nouvo Cimento 15 (1968) 695.

\bibitem{ramond} P.Ramond. Dual Theory for Free Fermions. 
Phys.Rev. D3 (1971) 2415
\bibitem{dirac1} P.A.M.Dirac. Proc.Roy.Soc. A155 (1936) 447;\\
M.Fierz and W.Pauli. Proc.Roy.Soc. A173 (1939) 211\\
W.Rarita and J.Schwinger. Phys.Rev. 60 (1941) 61\\
I.M.Gel'fand and M.A.Naimark. J.Phys. (USSR) 10 (1946) 93\\
Y.Nambu. Phys.Rev. 160 (1967) 1171\\
L.P.S.Sing and C.R.Hagen. Phys.Rev. D9 (1974) 910\\
J.Fang and C.Fronsdal. Phys.Rev. D18 (1978) 3630

\bibitem{15years}Superstrings. V1,V2. World Scientific 1985. (Ed. by J.Schwarz)


\bibitem{heisenberg}M.Born, W.Heisenberg and P.Jordan. Z.Physik 35 (1926) 557\\
H.Weyl. Math.Z. 24 (1926) 342\\
R.Brauer and H.Weyl. Amer.J.Math. 57 (1935) 425\\
E.Cartan. Bull.Soc.math.France 41 (1913) 53\\
Van der Waerden. Nachr.Ges.Wiss.Gott. (1929) 100\\
E.P.Wigner. Ann.Math. 40 (1939) 149\\
O.Laporte and G.Uhlenbeck. Phys.Rev. 37 (1931) 1380


\end{thebibliography}
\end{document}